# Web Spam Detection Using Multiple Kernels in Twin Support Vector Machine

Seyed Hamid Reza Mohammadi, Mohammad Ali Zare Chahooki

Yazd University, Yazd, Iran

mohammadi_6468@stu.yazd.ac.ir

chahooki@yazd.ac.ir

## ABSTRACT

Search engines are the most important tools for web data acquisition. Web pages are crawled and indexed by search Engines. Users typically locate useful web pages by querying a search engine. One of the challenges in search engines administration is spam pages which waste search engine resources. These pages by deception of search engine ranking algorithms try to be showed in the first page of results. There are many approaches to web spam pages detection such as measurement of HTML code style similarity, pages linguistic pattern analysis and machine learning algorithm on page content features. One of the famous algorithms has been used in machine learning approach is Support Vector Machine (SVM) classifier. Recently basic structure of SVM has been changed by new extensions to increase robustness and classification accuracy. In this paper we improved accuracy of web spam detection by using two nonlinear kernels into Twin SVM (TSVM) as an improved extension of SVM. The classifier ability to data separation has been increased by using two separated kernels for each class of data. Effectiveness of new proposed method has been experimented with two publicly used spam datasets called UK-2007 and UK-2006. Results show the effectiveness of proposed kernelized version of TSVM in web spam page detection.

## Indexing terms/Keywords

Search Engine, Web Spam Page Detection, Machine Learning, Twin Support Vector Machine (TSVM), Multiple Kernels.

**1. Introduction**

The ever-increasing volume of information on the Internet has made search engines vital tools for information extraction. From a theoretical point of view, search engines are information retrieval tools responsible for downloading, indexing, processing, retrieving, and ranking of information [1].

Malicious content on the Internet, poses a unique challenge for search engines; one that other information retrieval systems do not normally face. Malicious content refers to deceitful information created for the purpose of manipulating a search engine's results. The open nature of the Internet allows any individual to create and distribute arbitrary content. Thus, malicious content cannot be avoided. Web pages which attempt to manipulate a search engine's results are the most important type of malicious content. Such pages are called Spam Pages. Other types of malicious content include spam queries, spam posts and Fishing pages.

Most website owners want to have higher-ranking pages. This is because only 15% of users visit the second page of search engine results before changing their query. Thus, search engines are constantly trying to assign better ranks to the most relevant results. This improvement increases their revenue and popularity. Such attempts can be exploited, and led to the creation of spam pages by profiteers [2].

It should be noted that not all spam pages have commercial or exploitive purposes. The contents of these pages differ from the user's expectations. They aim to penetrate ranking algorithms and, therefore, can be classified as spam pages. The main purpose of spam pages is to fraud ranking algorithms, with the aim of achieving a higher rank among the top ten search engine's results of various queries, as fast as possible [3]. Fig.1 illustrates a sample spam page. Since the text in each page include multiple valuable keywords, the content of the page is insignificant for the user. From the perspective of search engine, the most important effect of spam pages is diminution of search engine user's trust. Furthermore, the large number of spam pages, forces search engines to allocate more resources (bandwidth, storage space, software) to crawling, indexing, storing, and processing contents of these pages.

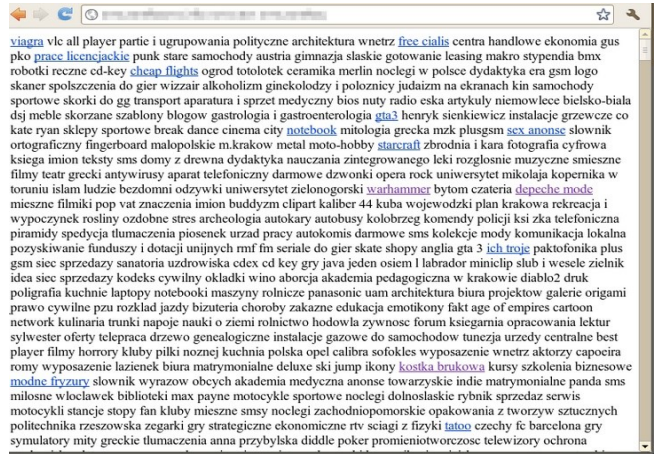

**Figure 1: An example spam page; although it contains popular keywords, the overall content is useless to a human user.**

Spam pages mainly aim to corrupt ranking algorithms in search engines. Ranking is done by search engine which estimates the quality of a page with respect to a particular query. In other words, one scores is assigned to each retrieved document. Documents are sorted in decreasing (or increasing) order of their scores. Search engines often use the following measures to rank web pages:

a) Relevance, measured as the degree of similarity between the input query and the text of the web page.

b) Importance, input query is ignored in it. This measure can be calculated as the number of times other pages are linked to a certain page.

As mentioned earlier, spam pages are created with the aim of obtaining a high rank in a search engine's results. As a result, they are created with the ranking algorithms in mind [1].

The content of a spam page is organized to match more queries. Therefore, a variety of useful keywords are used so that more user queries can be matched. Also, web spam content are changed based on the structure of web pages. Therefore, content ranking algorithms assign higher scores to these pages. If the goal is to match the target page to a certain request, then the target keywords need to be repeated multiple times. In 2006, Najork et al. [4] presented an approach based on content analysis, using examination of various features of spam and non-spam pages. Their findings formed a basis for identifying spam pages.

Creating a high number of links is another method used by spammers to improve their rank. In [6-8] a different approach, based on the links between pages in the web graph, is used for detection. The graph and patterns that emerge from it are used to differentiate spam and non-spam pages.

There have also been attempts to combine ideas form both methods to achieve satisfactory results. In [5], spam pages are detected using a combination of content-based and link-based methods. In recent years, user behavior has also been a decisive factor in detecting spam pages. Liu et al. [9] presented a behavior-based method. Modern techniques for creating spam pages include information hiding, where users are encountered with different versions of pages compared to search engine crawlers.

Various methods of spam detection have been proposed, including those based on coding-style similarity [18] and language pattern analysis [11]. Machine learning algorithms have also been applied to page features [12-16].

The majority of methods involved in creating spam pages rely on content-based techniques. Thus, analyzing the content features of these pages will improve much more efficiency compared to other methods. Web pages have certain features that can be extracted using content analysis. The values for these features are significantly different across spam and non-spam pages. Since the detection problem seeks to label pages as either *spam* or *normal*, it can be converted to a classification problem with two classes. Therefore, once features of page content are extracted and a suitable set of pages are labeled (normal or spam), a proper classifier can be used to determine the being spam of unlabeled pages collected by a search engine.

Machine learning methods use classification algorithms to extract and detect patterns among normal and spam pages. Using more powerful algorithms can increase the accuracy of detection. In previous researches, Hidden Markov Model [12], Decision tree [13], Bayesian networks [14], and artificial neural network [15] are used to separate spam instances from non-spam ones.

Many machine learning algorithms have been used for the binary classification problem. Support Vector Machine (SVM) is one of these algorithms, which offers favorable results. SVM despite is successfully applied to many problems of differentiating positive and negative instances, suitable accuracy has not been reported for web spam detection by it. However, studies on other applications demonstrate that new extensions of SVM have significantly superior performance [19]. Such versions aim to improve its performance on different data through changes in SVM structure.

SVM is binary classifier, which constructs a hyper-plane among instances such that the distance from the instances is maximized and, thus, a good separation is achieved. In an extended version of it, namely the Twin SVM (TWSVM), two separate hyper-planes are used for each class instances. Based on the nature of web pages and number of extracted features in them, this version of SVM can be used for web spam classification.

In some cases, samples cannot be separated linearly by their attributes. The concept of kernel has been introduced in machine learning methods to solve this problem. In this approach, the data are first logically mapped to a higher-dimensional space. The new dimensions are created such that they can be linearly separated with more accuracy, using hyper-planes [21].

In this paper, TSVM with two non-linear kernels is used for spam page detection. The main innovation of the proposed method is embedding two different suitable kernels for each hyperplane in TSVM. Experimental results on UK-2006 and UK-2007 datasets demonstrated the effectiveness of the proposed method for detecting spam pages.

The remainder of the paper is organized as follows. Section 2 presents a review of various extensions to the SVM. The importance of using non-linear kernels is considered in section 3. The details of the proposed method are discussed in section 4. Section 5 presents the experimental results. Finally, in section 6, the concluding remarks and suggestions for future works are offered.

**2- SVM Extensions**

Support Vector Machine (SVM) is a non-statistical binary classifier, which has attracted a lot of attentions in recent years. In this method, all the instances are used in conjunction with an optimization algorithm to find instances that form the boundaries of classes. Using these instances, an optimal linear decision boundary is created to separate classes, and it is solved using the Quadratic Programming Problem (QPP) [23]. This classifier does not rely on statistics; the training points are directly used to determine the decision boundary between two classes.

Proximal Support Vector Machine (PSVM) is a new version of the SVM, which aims to reduce computational costs. In this algorithm, two parallel hyper-planes, rather than one, are used to separate instances.

Generalized Eigenvalue Proximal SVM (GEPSVM), uses two non-parallel hyper-planes, each of which is placed near one of the two classes. Since a large QPP must be solved, the algorithm becomes very costly.

The improved GEPSVM, called Twin SVM (TWSVM), eliminates this problem by solving two smaller instances of QPP. Therefore, computational cost is decreased, while the algorithm becomes more robust.

Then, a variation of TWSVM, namely the Least Square TWSVM was introduced. The final solution of the algorithm is obtained by solving two linear equations, rather than two QPPs. Although the algorithm is easier, however it is less robust, due to its sensitivity to noise.

Knowledge-based TWSVM is another version of TWSVM, which uses current instances along with previous knowledge to separate instances into two classes. Compared to other knowledge-based algorithms, it offers less complexity time.

In order to improve the structure of TWSVM, Smooth TWSVM was introduced. The most important change in this version of the algorithm is the elimination of QPP limitations.

## 3- Non-linear Kernels

Using linear models for classification is much easier and faster than non-linear counterparts. Linear models work on data that can be separable linearly. Often, developing of linear models is more appropriate than non-linear ones, since they are associated with lower complexity time. However, in some cases, the instances are not intrinsically linear. Such instances must be mapped to a more dimensional space, where a linear model is applicable. Fig.2 illustrates how the data are mapped to a new space. Mapping to a high-dimensional space is not always straightforward manner and can impose much more costs than primary dimension space. By offering a way to reduce associated costs, the idea of *kernel* addresses this shortcoming. A kernel implicitly maps the data to a higher-dimensional space. In order to classify non-linear data in a linear manner, the SVM needs to use the idea of a kernel, which leads to enhanced classification capabilities [21].

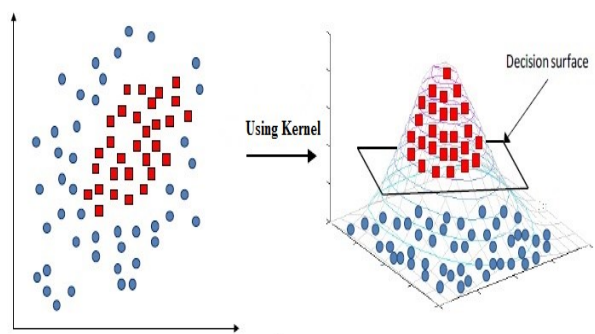

Figure 2 idea of using kernel

Generally, in algorithms where data are in the form of dot products, the idea of a kernel can be applied. There are various types of kernel functions. Polynomial, Gaussian, linear, and fuzzy are more common kernel functions [21]. Using kernel functions together with SVMs is lead to performance improvement of binary classification.

## 4- Proposed Method

In this paper, non-linear kernels in the extended SVM are used to present a binary classifier, with superior performance in detecting spam pages. In the following, the learning structure of the SVM is discussed in detail. Next, the non-linear kernels in the TWSVM are introduced. Finally, the proposed method, i.e. Multiple Kernel TSVM, is presented. As shown in section of experimental results, this method improves the accuracy of detecting spam pages.

## 4-1- TSVM

The TSVM was first introduced by Jayada [22], based on the normal SVM. In this algorithm, instead of using a single plane and increasing its margins toward binary classification, two non-parallel planes are used to separate instances into two classes. In this classifier, each class is determined using one of the hyper-planes. The procedure is demonstrated in Fig.3.

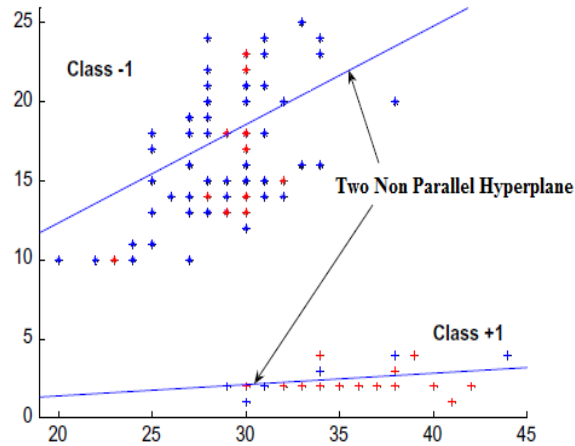

Figure 3: Operation of Twin Support Vector Machine (TSVM)

Assume that we are given a set of data containing positive and negative instances. In order to separate them, two equations are formed for each hyper-plane, which can segregate the data in TWSVM in a linear fashion.

$$F_+(X) = W_+^T X + b_+, \quad F_-(X) = W_-^T X + b_- \quad (1)$$

where $W_+, W_- \in \Re^n$ and $b_+, b_- \in \Re$

In the next step, two QPPs must be solved to obtain the optimal hyper-planes. QPPs can be challenging to solve; thus, they should be simplified and solved using Lagrange's method of undetermined coefficients. Once the QPP is solved, the final equation of the TWSVM to separate the instances is obtained in as follows.

$$Class = \arg \min_{i=1,2} \frac{|x^T w_i + b_i|}{\sqrt{w_i^T x^T w_i}} \quad (2)$$

where |.| is the absolute value and refers to classes of instances

Any new instance can be classified using Eq(2).

## 4-2- Using kernels in the TWSVM

As mentioned earlier, in order to convert non-linear classification to a linear one, instances are mapped to a new high-dimensional space before classification phase. In section 3, the idea of mapping the data to a new space was explained in detail. Furthermore, in the previous subsection, the final equation for the TWSVM classifier was presented. Here, we aim to rewrite this equation using a kernel function. Eq(3) shows the final classifier by kernel design.

$$Class = \arg \min_{i=1,2} \frac{|K(x^T, C^T)w_i + b_i|}{\sqrt{w_i^T K(x^T, C^T) w_i}} \qquad (3)$$

Where K is kernel function which maps data to higher dimension and any arbitrary kernel can be used to replace K, in this equation.

**4-3- Multiple non-linear kernels in the TWSVM**

As previously mentioned, in this paper for the first time two separate kernels are adjusted in TWSVM. The rationale behind this is the nature of spam pages as well as the distribution of instance. This paper tries to improve classification performance by designing two suitable kernels, both of which are used to map the input data to a new space. In the following, details of designing and matching two appropriate kernels for the TWSVM are discussed.

In subsection 4-2, kernel design in the TSVM is explained. Based on dimension space of samples, an efficient kernel must be designed to separate the samples both linearly and optimally, in the new space. The choice or design of the kernel function has a direct impact on the efficiency of the classifier algorithm. Kernel functions can be created as combinations of standard kernels. The resulting kernel must satisfy Mercer's condition (Mercer's theorem checks the validity of kernel functions). In this paper, a suitable kernel function for spam web pages is obtained using the method of *fixed rule* and a combination of two standard kernels. The first kernel is linear, which is powerful and has low computational costs. The linear kernel is shown in Eq(4).

$$K(X_i, X_j) = X_i \cdot X_j + b \qquad (4)$$

where $X_i, X_j \in \Re^n$ and $b \in \Re$

Linear kernel is most effective when the data are normally distributed. Since non-spam pages have a normal distribution, this kernel function is used to map these input instances to a new space.

The other kernel function is the hyperbolic tangent, which is used for non-linear patterns in the data. The function can be seen in Eq(5).

$$K(X_i, X_j) = tanh(k X_i \cdot X_j - b) \qquad (5)$$

where $X_i, X_j \in \Re^n$ and $k, b \in \Re$

Since each kernel function creates a similarity matrix, it must be designed such that the matrix for each instance of the two classes matches the real patterns in the data. Therefore, the choice of the function is very important. In this paper, the linear method is used, which offers simplicity while maintaining the advantages of both kernel functions. Eq(6) presents the linear combination of the two functions.

$$K = tanh(k \times X^{train} \times X^{t'} - b)^2 + (X^{train} \times X^{t'}) \qquad (6)$$

where $X^{train}$ is the set of positive instances (spam) and $X^t$ represents the entire set of data. After several experiments, the coefficients *k* and *b* were assigned 1 and 0, respectively. A function must satisfy Mercer's conditions before it can be considered a kernel function. The conditions are true for the proposed function.

After the appropriate kernel is chosen, it must be adapted to the TSVM. As shown earlier, in the TWSVM classifier the class of a new instance is determined using Eq(7).

$$Class = \arg \min_{i=1,2} \frac{|K(x^T, C^T)w_i + b_i|}{\sqrt{w_i^T K(x^T, C^T) w_i}} \qquad (7)$$

In the proposed method, Eq(6) is applied to spam pages, whereas Eq(4) is used for normal instances. As mentioned before, the TWSVM classifier uses two separate hyper-planes for each instance, which allows the mapping of the input data to a new space. As a result, the hyper-planes become more effective in segregating of instances.

After the designed kernel is adjusted to the TWSVM, the classifier must be trained. A portion of instances in each dataset is used for training purposes. The trained classifier can then determine which class each new instance belongs.

## 5- Experimental Results

In this section, we present the details of implementation and testing of the proposed algorithm, including description of datasets, evaluation criterion, and experimental results.

### 5-1- Dataset

In order to train, execute, and determine the accuracy of the proposed algorithm, UK-2006 and UK-2007 datasets were used. These data sets have been used in numerous studies on spam pages. UK-2006 contains around 11400 hosts from .uk domain where 61.75% of documents are labeled as normal, 22.08% as spam, and 16.16% as unknown [26]. UK-2007 holds 114529 hosts from .uk domain where 94% of them are labeled as normal and the remaining 6% are spam.

In this paper we ignore documents with *Unknown* label. The samples in two datasets are randomly divided into two parts, 75% as train and other 25% as test sets.

### 5-2- Evaluation

In order to present the algorithm's results and compare them to those of other algorithms, a standard criterion is necessary for evaluation. Accuracy is a measure of how many instances are classified correctly. It is popular in the context of deception detection. Compared to other measures, it is quite robust.

$$Accuracy = \frac{TP + TN}{P + N} \qquad (8)$$

In order to evaluate the accuracy of our classifier, we employed a technique known as ten-fold cross validation [4]. Ten-fold cross validation involves dividing the judged data set randomly into 10 equally-sized partitions, and performing 10 training/testing steps, where each step uses nine partitions to train the classifier and the remaining partition to test its effectiveness.

### 5-3- Comparison of the proposed algorithm to SVMs

Despite acceptable performance in classifying various types of data, the standard SVM algorithm has failed in the context of spam pages. The standard SVM was implemented using several kernels. A comparison of the results can be seen in Table 1. As evident, the choice of kernel does not lead to significant changes in algorithm's performance, which may be due to its own structure. The standard SVM only uses one separating plane; thus, it is not able to differentiate between relatively similar spam and normal instances. There are many similar instances with opposing labels in the data sets. As a result, the SVM classifier demonstrates poor performance in web spam detection.

Table 1: Comparison between proposed method with SVM and TWSVM by different kernels

| Algorithm | Kernel | (٪)Accuracy UK-2006 | (٪)Accuracy UK-2007 |
|---|---|---|---|
| SVM | Linear | 82 | 80 |
| | RBF | 84 | 81.4 |
| | Hyperbolic | 84 | 84 |
| TWSVM (One Kernel) | Linear | 88 | 92 |
| | RBF | 88.3 | 92 |
| | Custom Kernel | 90 | 94 |
| TWSVM (Two Kernel) | Kernel1 = Linear Kernel2 = RBF | 90 | 92 |
| | Kernel1 = RBF Kernel2 = CK | 92 | 94 |
| | Kernel1 = Linear Kernel2 = CK | **93** | **95.6** |

## 5-4- Comparison to other algorithms

As mentioned before, different classifiers have been used for web sapm detection, including decision trees [14] and artificial neural networks [16]. Furthermore, other studies such as [13] and [15] use different algorithms, which cannot be compared to the proposed algorithm due to incompatible datasets. Tables 2 and 3 present our results along with those of the most accurate algorithms in KU-2006, UK-2007 respectively.

Table 2: Comparison between proposed method with other algorithm

| UK-2006 | | |
|---|---|---|
| Algorithm | (٪)Accuracy | F-M |
| Neural Network [16] | 87.6 | - |
| HMM [13] | - | 0.859 |
| Decision Tree[14] | - | 0.92 |
| TWSVM | 90 | .918 |
| MKTWSVM | **93** | **0.941** |

Table 3: Comparison between proposed method with other algorithm

| UK-2007 | |
|---|---|
| Algorithm | (٪)Accuracy |
| Genetic Algorithm [24] | 92.41 |
| Supervised Neural Network [25] | 93.66 |
| HTWSVM | 94 |
| MKTWSVM | 95.6 |

MKTWSVM which uses two hyper-planes to separate instances of the two classes, with two kernels for each, is able to achieve acceptable accuracy. Interestingly, in this method, the input data are mapped such that a linear model can be used for differentiation. The choice and design of the kernel is an important issue in this problem. The proposed kernel in this paper improves the accuracy of a linear classifier. Thus, a non-linear kernel can be considered the key component of a classifier. As shown in Table 3, nearly 3% of the increase in accuracy is associated with two separate kernels in the structure of the TWSVM. Advantages of this method include low computational complexity since the algorithm works in linear space.

**6- Conclusion and Future Works**

In this paper, by using two non-linear kernels in TSVM, high accuracy in detecting spam pages was achieved. Furthermore, in order to demonstrate the effectiveness of non-linear kernels, SVM and TWSVM were tested both with and without kernel functions. The proposed method proved that two non-linear kernels in TWSVM increases accuracy and reduces computational complexity.

An avenue for future studies involves altering the structure of extended versions of SVM so that accuracy is increased and computational complexity becomes more satisfactory. Moreover, since the optimal combination of basic kernel functions for either of the hyper-planes can improve classification accuracy, genetic programming can be incorporated to find the optimal combination of basic kernel functions.